\documentclass[RNAAS]{aastex62}

\begin{document}

\title{The nearest extreme velocity stars among Gaia DR2 high proper motion stars}

\correspondingauthor{Ralf-Dieter Scholz}
\email{rdscholz@aip.de}

\author[0000-0002-0894-9187]{Ralf-Dieter Scholz}
\affiliation{Leibniz Institute for Astrophysics Potsdam (AIP),\\
An der Sternwarte 16, 14482 Potsdam, Germany}

\keywords{parallaxes --- 
proper motions --- stars: kinematics and dynamics --- 
Galaxy: halo -- solar neighborhood}

\section{} 

The second data release (DR2) of the European Space Agency mission
Gaia \citep{2018A&A...616A...1G} has already been used in the
search for new extreme velocity stars that can be unbound to the Galaxy. 
Some of these investigations \citep[e.g.][]{2018MNRAS.tmp.2466M,
2018arXiv180503194H}
concentrated on the subsample of about 7 million
stars with both astrometric and radial velocity (RV) measurements in
Gaia DR2.
However, \citet{2018ApJ...865...15S} and \citet{2018arXiv180802620B} 
have shown that based on
Gaia DR2 parallaxes and proper motions alone one
can efficiently select relatively nearby high-speed candidates. 
The aim of this note is to check the role of various astrometric 
quality criteria in the selection of 
such candidates among Gaia DR2 high proper motion (HPM) stars.

In their search for hypervelocity runaway white dwarfs (HVR WDs), \citet{2018ApJ...865...15S}
investigated HPM objects with $\mu>211$\,mas/yr
and moderately significant parallaxes ($\varpi>3\sigma_{\varpi}$).
There are about 61,500 Gaia DR2 stars with these conditions,
slightly more than the number of classical HPM
stars in the New Luyten Two Tenths
\citep[NLTT;][]{1995yCat.1098....0L} catalog.
With a lower limit of 60\,mas/yr, three times smaller
than in the NLTT, we found about 1.8 million stars with $\varpi>3\sigma_{\varpi}$
in Gaia DR2. Following Eq.~1 of \citet{2018arXiv180503194H} but preferring the
local standard of rest velocity of 235\,km/s (instead of 220\,km/s) used by
\citet{2018arXiv180802620B}, we computed the Galacic rest frame
tangential velocities $v_{t,g}$ of the stars corrected for solar motion.
By simply applying an arbitrary lower limit of $v_{t,g}$$>$700\,km/s,
well above the local Galactic escape velocity of 580$\pm$63\,km/s \citep{2018A&A...616L...9M},
we found 109 objects. Using now the Gaia DR2 quality criteria of
\citet{2018A&A...616A...2L}, concerning the astrometric unit weight error $u$
and $BP$/$RP$ photometry (critical in crowded fields), and the
recommendations in the basic source parameter descriptions
\citep{2018A&A...616A...1G} \textsf{astrometric\_gof\_al}$<$3 and
\textsf{astrometric\_excess\_noise\_sig}$\le$2, also applied
by \citet{2018MNRAS.tmp.2466M} and \citet{2018arXiv180802620B}, only
39 candidates (with parallaxes between 0.17\,mas and 1.24\,mas)
remained. All three HVR WDs of \citet{2018ApJ...865...15S}
but no additional such candidates (falling in the same region of the
colour-magnitude diagram; see their Fig.10) are among them.

In the following, we considered only objects with highly significant parallaxes ($\varpi>5\sigma_{\varpi}$)
\citep[cf.][]{2018arXiv180802620B} as reliable extreme tangential velocity candidates.
Out of the selected 39 objects, there are six such high-priority candidates.
Their main photometric and astrometric Gaia DR2 \citep{2018A&A...616A...1G} data 
are listed in Table~\ref{tab:1} (sorted by decreasing parallax), 
including the astrometric unit weight error 
$u$ \citep{2018A&A...616A...2L} and other astrometric quality parameters.
Among them, we list the \textsf{visibility\_periods\_used}, for
which \citet{2018MNRAS.tmp.2466M} and \citet{2018arXiv180802620B} used minimum 
numbers of 9 and 10, respectively, whereas we did not select a lower limit.
While two of the three HVR WDs presented by \citet{2018ApJ...865...15S},
D6-1 (= new HPM star Gaia DR2 5805243926609660032) and D6-3 (= LSPM J1852+6202), 
were excluded because of their only moderately significant
parallaxes, the third one, D6-2 (= NLTT 51732), is
the nearest among all candidates in Table~\ref{tab:1} and has 
one of the largest $v_{t,g}$. However, its quality parameters $u$, 
\textsf{astrometric\_gof\_al}, and \textsf{astrometric\_excess\_noise\_sig}
are close to the maximum allowed values, and the number of \textsf{visibility\_periods\_used}
is relatively low. Therefore, and because of the zero RV 
measured by \citet{2018ApJ...865...15S}, we consider this object, the only 
classical HPM star in our list, a doubtful candidate. On the
other hand, Gaia DR2 1540013339194597376, the nearest 
high tangential velocity (and HPM) star found by \citet{2018arXiv180802620B},
is probably the best high-speed star candidate (small astrometric quality parameters
and high number of \textsf{visibility\_periods\_used}) in our list. Our most distant
candidate, Gaia DR2 6097052289696317952, has the highest $v_{t,g}$ and
appears slightly questionable with respect to its low number of \textsf{visibility\_periods\_used},
but interesting with its bright $G$ magnitude (corresponding to an absolute magnitude of
about $-$0.3) and relatively blue $BP$$-$$RP$ colour. Further RV measurements
will help to clarify the high-speed and possibly unbound status of our candidates.


\begin{deluxetable}{ccccrrccrrc}
\tablecaption{Selected Gaia DR2 stars with proper motions $\mu$$>$60\,mas/yr and  Galactocentric tangential velocities $v_{t,g}$$>$700\,km/s\label{tab:1}}
\tablehead{
\colhead{Gaia DR2 ID} & \colhead{$G$} & \colhead{$BP$$-$$RP$} & \colhead{$\varpi$} & \colhead{$\mu_{\alpha}\cos{\delta}$} & \colhead{$\mu_{\delta}$} & \colhead{$u^1$} & \colhead{$D^2$} & \colhead{gofAL$^3$} & \colhead{vpu$^4$} & $v_{t,g}$ \\             
                      & [mag]         & [mag]                 & [mas]              & [mas/yr]                             & [mas/yr]                 &               &               &                 &               & [km/s]
}
\startdata
1798008584396457088$^5$ & 17.02 & 0.40 & 1.05$\pm$0.11 &  $+$98.39$\pm$0.21 & $+$240.35$\pm$0.17 & 1.13 &  1.8 &    2.7 &  9 & 1248$\pm$122 \\
1540013339194597376$^6$ & 15.96 & 1.01 & 0.59$\pm$0.05 &  $-$82.03$\pm$0.05 & $-$118.29$\pm$0.05 & 0.98 &  0.0 & $-$0.4 & 15 &  917$\pm$104 \\
6698855754225352192$^6$ & 15.39 & 0.81 & 0.47$\pm$0.05 &  $-$38.97$\pm$0.07 &  $-$86.57$\pm$0.05 & 1.02 &  0.0 &    0.3 & 13 &  733$\pm$99 \\
3841458366321558656     & 15.89 & 0.86 & 0.33$\pm$0.06 &   $+$7.29$\pm$0.11 &  $-$81.39$\pm$0.11 & 0.92 &  0.0 & $-$1.0 &  8 &  978$\pm$220 \\
3593446274383096448     & 14.03 & 0.89 & 0.27$\pm$0.04 &  $-$36.35$\pm$0.06 &  $-$47.95$\pm$0.04 & 0.92 &  0.0 & $-$1.2 &  9 &  855$\pm$156 \\
6097052289696317952     & 13.53 & 0.60 & 0.17$\pm$0.03 &  $-$61.10$\pm$0.05 &  $-$24.73$\pm$0.05 & 1.02 &  0.0 &    0.3 &  9 & 1617$\pm$347 \\
\enddata
\tablecomments{$^1$unit weight error, $^2$\textsf{astrometric\_excess\_noise\_sig}, $^3$\textsf{astrometric\_gof\_al}, $^4$\textsf{visibility\_periods\_used}, $^5$(= NLTT 51732) classified as hypervelocity white dwarf D6-2 \citep{2018ApJ...865...15S}, $^6$previously found nearby high-speed star candidate \citep{2018arXiv180802620B}}
\end{deluxetable}


\begin{thebibliography}{}

\bibitem[Bromley et al.(2018)]{2018arXiv180802620B} Bromley, B.~C., Kenyon, S.~J., Brown, W.~R., \& Geller, M.~J.\ 2018, arXiv:1808.02620

\bibitem[Gaia Collaboration et al.(2018)]{2018A&A...616A...1G} Gaia Collaboration, Brown, A.~G.~A., Vallenari, A., et al.\ 2018, \aap, 616, A1

\bibitem[Hattori et al.(2018)]{2018arXiv180503194H} Hattori, K., Valluri, M., Bell, E.~F., \& Roederer, I.~U.\ 2018, arXiv:1805.03194

\bibitem[Lindegren et al.(2018)]{2018A&A...616A...2L} Lindegren, L., Hern{\'a}ndez, J., Bombrun, A., et al.\ 2018, \aap, 616, A2

\bibitem[Luyten(1995)]{1995yCat.1098....0L} Luyten, W.~J.\ 1995, VizieR Online Data Catalog, 1098,

\bibitem[Marchetti et al.(2018)]{2018MNRAS.tmp.2466M} Marchetti, T., Rossi, E.~M., \& Brown, A.~G.~A.\ 2018, \mnras,

\bibitem[Monari et al.(2018)]{2018A&A...616L...9M} Monari, G., Famaey, B., Carrillo, I., et al.\ 2018, \aap, 616, L9

\bibitem[Shen et al.(2018)]{2018ApJ...865...15S} Shen, K.~J., Boubert, D., G{\"a}nsicke, B.~T., et al.\ 2018, \apj, 865, 15

\end{thebibliography}
\end{document}